\documentclass[12pt]{spieman}  % 12pt font required by SPIE;

\usepackage[utf8]{inputenc}
\usepackage[T1]{fontenc}
\usepackage{amsmath,amsfonts,amssymb}
\usepackage{graphicx}
\usepackage{setspace}
\usepackage{tocloft}
\usepackage{lineno}
\usepackage{astro_bib_macro}
\usepackage{xspace}
\usepackage{soul}

\title{Impact of segmented deformable mirrors on high-contrast testbeds for exoplanet imaging with future large space telescopes: contrast stability assessment on the HiCAT bench}

\author[a,*]{Benjamin Buralli}
\author[a]{Mamadou N'Diaye}
\author[b,c]{Raphaël Pourcelot}
\author[a]{Marcel Carbillet}
\author[d]{Emiel H. Por}
\author[a,e]{Iva Laginja}
\author[f]{Ludovic Canas}
\author[b]{Sarah Steiger}
\author[b,g]{Peter Petrone}
\author[k]{Meiji M. Nguyen}
\author[b]{Bryony Nickson}
\author[h,i]{Susan F. Redmond}
\author[j]{Ananya Sahoo}
\author[b]{Laurent Pueyo}
\author[b]{Marshall D. Perrin}
\author[b]{Rémi Soummer}
\affil[a]{Université Côte d’Azur, Observatoire de la Côte d’Azur, CNRS, Laboratoire Lagrange, Nice, France}
\affil[b]{Space Telescope Science Institute, Baltimore, USA}
\affil[c]{Max Planck Institute for Astronomy, Heidelberg, Germany}
\affil[d]{University of California, Santa Cruz, USA}
\affil[e]{LIRA, Observatoire de Paris, Meudon, France}
\affil[f]{Thales Alenia Space, Cannes, France}
\affil[g]{NASA Goddard Space Flight Center, Greenbelt, USA}
\affil[h]{Jet Propulsion Laboratory, California Institute of Technology, 4800 Oak Grove Dr, Pasadena, CA 91011, USA}
\affil[i]{California Institute of Technology, 1200 East California Boulevard, Pasadena, CA 91125, USA}
\affil[j]{LoCSST, University of Massachusetts Lowell,  Lowell, Massachusetts, USA}
\affil[k]{Johns Hopkins University, Baltimore, Maryland, USA}

\cftpagenumbersoff{figure}
\cftpagenumbersoff{table} 
\begin{document} 
\maketitle

\begin{abstract} % 200 words

We investigate the stability of a segmented deformable mirror (DM) on high-contrast testbeds and its impact on the images produced with coronagraphs. Segmented apertures are promising to obtain large primary mirrors for future missions with starlight suppression capabilities. Cophased at the sub-nanometer level, segments can be slightly misaligned by small drifts, proving harmful for exoplanet observations. We study the impact of misalignments on contrast using the High-contrast Imager for Complex Aperture Telescopes (HiCAT), a testbed which includes a 37-segment DM and produces coronagraphic images with $2.5\times 10^{-8}$ contrast in narrowband light. Temporal wavefront errors due to the segmented DM are estimated with a Zernike wavefront sensor. Our in-lab results show aberrations at the sub-nanometer level, proving encouraging for contrast stability studies. We then use a digital twin of HiCAT to simulate coronagraphic images with an initial $0.5 \times 10^{-8}$ contrast and the segments in flat position. By injecting known perturbations on the segments, we observe a contrast degradation by a factor of 2.5, nearly corresponding to the typical contrast observed on HiCAT. These results highlight the importance of segment cophasing sensing and control strategies to ensure the required contrasts for exo-Earth imaging with a large segmented aperture for the Habitable Worlds Observatory mission.

\end{abstract}

% Include a list of up to six keywords after the abstract

{\noindent \footnotesize\textbf{*}Benjamin Buralli,  \linkable{benjamin.buralli@oca.eu} }

\begin{spacing}{1} %{2}   % use double spacing for rest of manuscript

\section{Introduction}
\label{sect:intro} 

The detection and characterization of Earth-like exoplanets is one of the main objectives of modern astrophysics, as it provides a promising path to study planetary formation, evolution, and the possible emergence of life. Among the different observational techniques, direct imaging provides access to the orbital and physical parameters of such planets, as well as the opportunity to analyze their atmospheres through spectroscopy in the search for possible biosignatures. However, imaging Earth-like exoplanets proves extremely challenging: it requires disentangling the faint signal of an exo-Earth from its host Sun-like star at contrast levels of the order of $10^{-10}$ with sub-arcsecond angular separations \cite{Trauger2007}. One promising approach to pursue such levels of starlight suppression is through the combination of wavefront control and coronagraphy, which aims to attenuate the light of an observed star to reveal its close-in faint companions. On the ground, atmospheric turbulence represents a challenging limitation to overcome and achieve such contrast levels. Solutions combining coronagraphs and extreme adaptive optics (XAO) \cite{Guyon2018} are currently investigated to reach contrasts down to $10^{-9}$ and observe Earth analogs around M dwarfs with Extremely Large Telescopes \cite{Kasper2021}. In space, solutions integrating coronagraphs and active wavefront control \cite{Por_2023, Wallace_2025} appear as a suitable path with large ultra stable observatories towards $10^{-10}$ contrast and the detection of Earth-like planets around Sun-like stars.

Large-aperture space telescopes, such as the James Webb Space Telescope (JWST) \cite{JWST_2023, JWST_2024}, rely on segmented primary mirrors deployed and aligned to work as a monolithic optical surface \cite{Feinberg_2022, Lajoie_2023}. While this architecture enables much larger collecting areas than monolithic aperture telescopes and avoids atmospheric perturbations, it also introduces new challenges. In particular, thermo-mechanical drifts can produce nanometer-scale segment misalignments, generating mid-order spatial frequency aberrations that degrade coronagraphic performance and limit the achievable contrast down to $10^{-6}$ \cite{Laginja2021,Ruane2016}. Maintaining a stable contrast therefore requires precise phasing control and real-time wavefront correction. These control strategies need to be validated experimentally on optical testbeds using segmented deformable mirrors (DMs), which both simulate segment drifts and provide active correction \cite{Helmbrecht2016, JaninPotiron2017}.

High-contrast testbeds are critical for developing and validating system-level approaches to direct imaging of exo-Earths with large segmented aperture telescopes. These facilities generate images at high contrast and allow precise characterization of hardware limitations, informing the design of future missions. Several testbeds in the world employ segmented deformable mirrors to study segment alignment, wavefront control, or coronagraphic performance. Examples include the Segmented Pupil Experiment for Exoplanet Detection (SPEED) testbed \cite{Beaulieu_2018, Beaulieu_2020}, the Santa Cruz Extreme Adaptive optics Lab (SEAL) \cite{Jensen-Clem_2021, Moreno_2024}, and Caltech’s High-Contrast Spectroscopy Testbed (HCST) \cite{Jovanovic_2018, Bertrou-Cantou_2024}. The High-contrast imager for Complex Aperture Telescopes (HiCAT) \cite{Soummer_2024} produces images reaching contrast levels of approximately $10^{-8}$ with an IrisAO segmented DM composed of 37 hexagonal segments. Its design allows detailed investigation of segmented mirrors stability and quantification of their impact on achievable contrast, providing critical insights to guide the development of next-generation high-contrast imaging testbeds \cite{Soummer_2024}.

Beyond hardware development, these studies are also critical for preparing future flagship missions such as the Habitable Worlds Observatory (HWO). Recommended by the Astro2020 Decadal Survey \cite{Decadal_survey_2021}, HWO will operate as an observatory for general astrophysics purposes. In addition, it aims to image and spectrally characterize Earth analogs, requiring starlight suppression down to flux ratios of the order of $10^{-10}$ \cite{Decadal_survey_2021, NASA_HWO_2024}. Investigating the performance limitations imposed by segmented mirrors and their stability provides the specifications needed for future high-contrast testbeds and informs the architecture of HWO-class missions. Previous work using the Pair-based Analytical model for Segmented Telescopes Imaging from Space (PASTIS) model has not only highlighted the sensitivity of coronagraphic contrast to segment-level aberrations, but also provided a practical framework for deriving segment error budgets by relating local aberrations to dark-hole contrast degradation \cite{Leboulleux2020, PASTIS2018}.

In this context, the relevance of studying the temporal stability of a segmented DM should be interpreted with respect to the stability requirements of large segmented space telescopes. While the primary mirror of such observatories is governed by its own structural dynamics and dedicated phasing control system, segmented DMs play a complementary role within the optical train of an instrument and in high-contrast laboratory testbeds. These devices can both be used to (i) correct for the residual segment-level errors left after primary mirror phasing and (ii) emulate piston, tip, and tilt perturbations representative of realistic segment drifts in a controlled and repeatable manner.

Rather than trying to replicate the dynamics of a full-scale primary mirror, characterizing the intrinsic stability and temporal behavior of a segmented DM aims to quantify how small, slowly evolving segment-level mid-order aberrations propagate into coronagraphic contrast degradation \cite{Laginja_2022}. Such experimental measurements provide a practical link between laboratory demonstrations and mission-level requirements, enabling the validation of wavefront control strategies and the derivation of segment-level stability constraints relevant to future large segmented aperture space telescopes. In addition, detailed knowledge of the segmented DM stability can inform the setup, calibration, and interpretation of high-contrast testbed experiments, helping to distinguish intrinsic DM drifts from other sources of wavefront errors and to assess their impact on achievable contrast.

At a system-level approach, quantifying the limitations to contrast performance stemming from individual components is essential for improving testbed results and developing analytical tools. HiCAT is a laboratory experiment designed to demonstrate coronagraphy with a segmented aperture, providing a platform for such investigations. Advancing the understanding of segmented aperture behavior is critical for future missions, which will rely on these tools to achieve their scientific objectives.

This paper presents an experimental study focused on the temporal stability of a segmented DM and its impact on coronagraphic contrast. Our objective is to determine the stability of wavefront errors introduced by the DM, to model their effect on the current $10^{-8}$ contrast regime measured on HiCAT, and to identify the limitations imposed by this device. We describe our experimental setup, including the integration of a Zernike-based mid-order wavefront sensor (MOWFS), present results from the stability analysis, and provide simulation-based estimates of achievable contrast limits. Finally, we discuss the implications of these results for future testbeds and HWO-class missions \cite{Buralli_2024, Soummer_2024}.

\section{Experimental setup}

We present the results that have been obtained on the HiCAT testbed, which includes an IrisAO segmented DM. Our objective is to provide insights into the usefulness of quantifying testbed segmented DM stability, and more specifically in the context of high-contrast imaging. 

\subsection{HiCAT}

\begin{figure}[ht!]
    \centering
    \includegraphics[width=1\columnwidth]{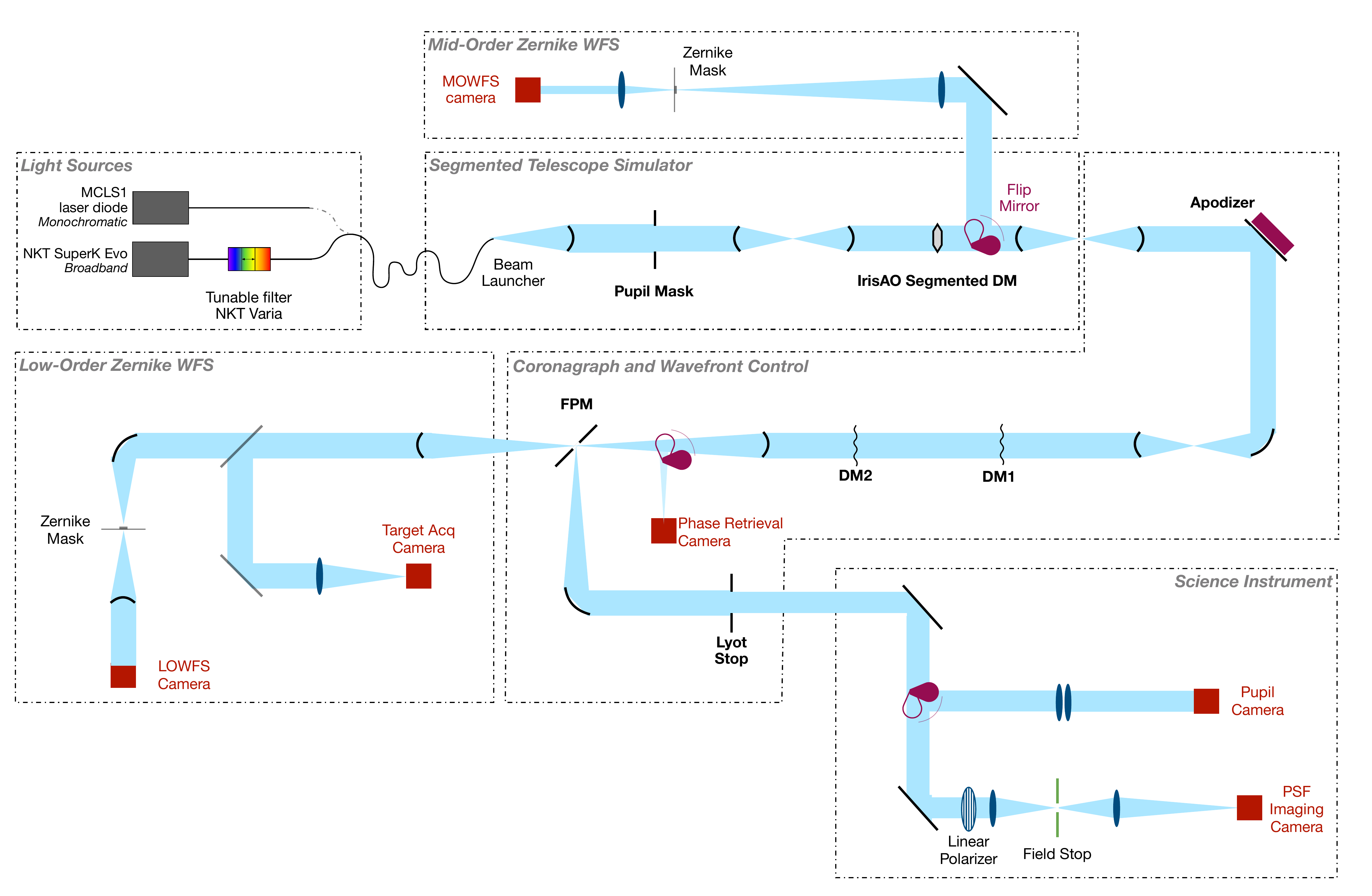}
    \caption{HiCAT system diagram as of the first semester of 2025. The schematic is divided into functional blocks. it is not to scale, and does not depict the physical arrangement of components on the table. For clarity, many optics which are depicted as transmissive are actually reflective.}
    \label{fig:schematic}
\end{figure}

HiCAT \cite{NDiaye_2012+1_2, NDiaye_2014, NDiaye_2015_2, Leboulleux_2016, Leboulleux_2017, HiCAT_Software, Soummer_2018, Soummer_2024} is a high-contrast imaging testbed designed to mature coronagraphy and wavefront control technologies for future segmented-aperture space telescopes. The system is organized into functional blocks, with each block corresponding to a key stage of the optical path, see Figure~\ref{fig:schematic}. In the ``Light Sources" block, illumination is provided by a monochromatic laser diode or a broadband supercontinuum source, combined with a tunable filter for spectral and intensity control. The beam then propagates through the coronagraph and wavefront control section, as represented in ``Segmented Telescope Simulator". In the following block, ``Coronagraph and Wavefront Control", an apodizer shapes the pupil amplitude, followed by two continuous deformable mirrors DM1 and DM2 located at the pupil and out of pupil for active phase and amplitude correction. A focal plane mask (FPM) blocks the light of the on-axis source, while a Lyot stop removes the diffracted light by the FPM. A phase retrieval camera measures residual wavefront errors and supports iterative phase reconstruction for fine calibration. Together, these components enable precise wavefront shaping and starlight suppression. In the ``Low-Order Zernike WFS", low-order aberrations and pointing errors are monitored by a Zernike wavefront sensor (ZWFS), providing feedback for tip–tilt stabilization and low-order correction \cite{Pourcelot_2022, Pourcelot_2023}. Similarly, in the ``Mid-Order Zernike WFS" block, the mid-spatial frequency contents of the aberrations are measured by the MOWFS, which employs its own Zernike wavefront sensor (WFS) placed after the IrisAO DM. The ``Science Instrument" module includes cameras for pupil imaging and point-spread function (PSF) acquisition, providing final science measurements. 

HiCAT can perform coronagraphy in several configurations: Classical Lyot Coronagraph (CLC), Apodized Pupil Lyot Coronagraph (APLC)\cite{Soummer_2005, NDiaye_2015, NDiaye_2016b, Zimmerman_2016}, and Phase-Apodized Pupil Lyot Coronagraph (PAPLC)\cite{Por_2020}. Its software package, shared among the HiCAT collaborators, consists of a set of scripts and parameter files built on the \texttt{CATKit2} library \cite{Por_2024}. This package allows for the control of both the HiCAT testbed and its digital twin, known as the HiCAT simulator \cite{HiCAT_Software}. The software also implements an electric field conjugation (EFC) algorithm \cite{Borde_2006, Giveon_2007, Will_2023}, which uses two continuous Boston Micromachines Corp. (BMC) DMs to generate a high-contrast dark zone in the coronagraphic image.

After the light is sent to the testbed through an optical fiber, an outline frame pupil mask defines the edges of the telescope primary mirror, followed by an IrisAO PTT111L segmented DM with 37 segments, distributed over three rings around a central segment. Each segment has a point-to-point diameter of 1.4 mm, forming a pupil with a circumscribed diameter of 7 mm. Under each segment are three micro-electro-mechanical systems (MEMS) actuators to generate piston, tip, and tilt modes. A 14-bit electronic controller drives the system, and each actuator has a maximum stroke of 5\,$\mu$m. 

The results published in Soummer et al., 2024 \cite{Soummer_2024}, hereafter denoted 2024 HiCAT paper, show contrast levels of a few $10^{-8}$ in narrowband (3\% bandpass) and broadband (9\% bandpass). For our study, we worked both in monochromatic light at 638\,nm and in narrowband at 680\,nm with a 3\% bandpass.

\subsection{MOWFS}

The MOWFS was developed to correct for the aberrations caused by the telescope segmented aperture. However, in its current implementation, the pick-off mirror directing the light to the MOWFS setup exhibited instabilities that affected the beam alignment. At the time of the study, it was not possible to alternate between the MOWFS setup and the science setup for comparison.  

\begin{figure}[ht!]
    \centering
    \includegraphics[width=0.6\columnwidth]{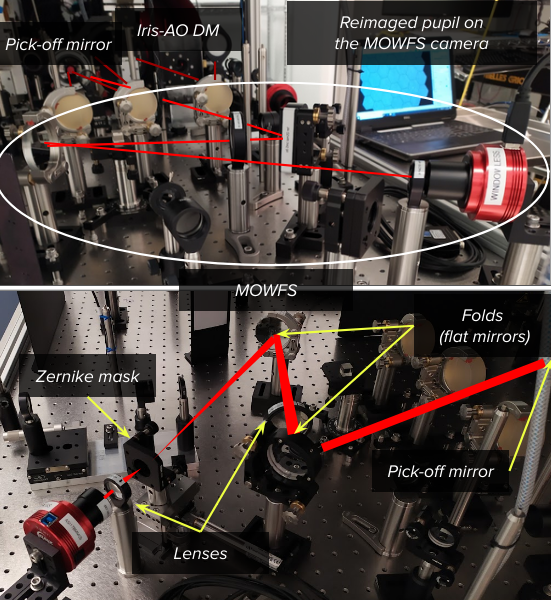}
    \caption{HiCAT mid-order wavefront sensor (MOWFS) setup. Both pictures show different views of the MOWFS. The top view shows it from up-front, with the rest of the HiCAT bench behind. On the right, the entrance pupil is re-imaged on the MOWFS detector and displayed on the laptop screen. The bottom view shows the MOWFS setup from high-up and the light beam is now coming from the right.}
    \label{fig:MOWFS}
\end{figure}

To assess the stability of the IrisAO DM segments, a wavefront control setup capable of measuring mid-order aberrations was required. The MOWFS was first integrated into HiCAT in June 2023, see Figure~\ref{fig:MOWFS}. A pick-off mirror was installed between the IrisAO DM and the apodizer to direct the light toward either  the coronagraphic path or the MOWFS setup. 
The MOWFS is a ZWFS specifically designed to measure DM segment misalignments, or mid-order aberrations. It consists of a series of lenses that form an image of the source on the ZWFS mask, followed by a CMOS camera positioned in the pupil plane to record the signal.

The ZWFS is based on the phase contrast method first developed for microscopy by Frits Zernike \cite{Zernike_1934}, and later adapted to astronomy \cite{Dicke_1975}. It has since found several applications such as (i) high-contrast imaging, e.g. calibration of non-common path aberrations to improve exoplanet imaging \cite{NDiaye_2012+1, NDiaye_2016, Vigan_2011, Vigan_2019, Ruane_2020}, (ii) adaptive optics, e.g. as a highly sensitive wavefront sensor or for instrumental developments \cite{Bloemhof_2003, Dohlen_2004, Wallace_2011, Jensen-Clem_2012}, (iii) segment phasing \cite{Surdej_2010, Jackson_2016, Janin-Potiron_2017, Van_Kooten_2022, Salama_2024} and (iv) picometer-level metrology for extremely sensitive wavefront error measurements \cite{Steeves_2020}.

The ZWFS uses a focal plane mask with a small central dot of up to a few resolution elements in diameter, designed to introduce a $\pi/2$ phase shift on the PSF core, so that the part of the electric field which is shifted interferes with the rest of the unaltered electric field, converting phase aberrations into intensity variations on the camera. The dot depth and diameter can be adjusted to optimize the sensor sensitivity \cite{Chambouleyron_2021}. 

The ZWFS was initially proposed with a relative diameter of 1.06 $\lambda/D$ \cite{NDiaye_2012+1}, with $\lambda$ and $D$ denoting the wavelength and the pupil diameter. It has been shown that a larger diameter would increase the sensitivity for higher order modes to the detriment of the lower ones, like global tip and tilt \cite{Chambouleyron_2021}. With a mask of 41\,$\mu$m of diameter at our disposal and the testbed room constraints on the optical layout, we selected optics leading to a beam focal ratio corresponding to a mask with an effective relative diameter of 2.5\,$\lambda/D$, providing a good balance between segment mode sensitivity and global tip-tilt sensitivity.
The main advantages of this type of wavefront sensor are its very high sensitivity to small aberrations amplitudes and its ease of implementation. However, the ZWFS is known for having a moderate capture range. In our study, this is not an issue as we are working in the small aberration regime (phase aberrations $\ll$ 1\,rad). In this regime, a linear relationship can be derived between the intensity variations on the detector and the phase aberrations, which can be very useful for wavefront error reconstruction. We can consider two methods of reconstruction, the analytical method and the interaction matrix method. The analytical method makes use of the exact analytical expression between intensity on the ZWFS camera and wavefront phase \cite{Pourcelot_2022PhD,Chambouleyron_2023, Wallace_2023}. This expression can be linearized in the small aberration regime \cite{NDiaye_2012+1}. The interaction matrix method relies on computing a matrix that describes the response of the wavefront sensor to commands applied to the IrisAO DM. In this paper, this method is used to reconstruct segment-level aberrations from the MOWFS measurements.

\section{Segmented DM stability with the MOWFS}\label{sec:DM_stability}

In this section, we present the results with the MOWFS installed on HiCAT and the interaction matrix reconstruction method to estimate the IrisAO segmented DM temporal stability.

\subsection{Experimental procedure}

To evaluate the temporal stability of the IrisAO segmented DM, we first establish a systematic experimental procedure which includes the calibration of the interaction matrix, the computation of the command matrix, and the reconstruction of segment aberrations from the MOWFS measurements. The interaction matrix is calibrated using the IrisAO DM in flat position, which was characterized in open loop with a Fizeau interferometer, showing an optical surface error of 9\,nm root mean square (RMS) \cite{Soummer_2018}.

To build the interaction matrix, also referred as Jacobian matrix $\mathbf{J}$, we decided to use piston, tip, and tilt (PTT) modes for each segment. This choice is driven by the DM architecture, as each segment is controlled by three actuators, allowing only PTT motions. Higher-order modes cannot be generated independently on individual segments. Normalized PTT modes are therefore applied sequentially to each of the 37 segments with an absolute amplitude $A_{abs}$ of 10\,nm peak-to-valley (PTV) and the corresponding image in the relayed pupil plane is recorded with the MOWFS CMOS camera. The value of $A_{abs}$ lies within the linear range of the ZWFS and is therefore used for all subsequent acquisitions.

To improve the calibration precision, both positive and negative amplitudes of the same $A_{abs}$ are injected. The resulting MOWFS intensities $\mathbf{J}_q^+$ and $\mathbf{J}_q^-$ are averaged and stored into a vector $\mathbf{J}_q$, which can be written as $\mathbf{J}_q = (\mathbf{J}_q^+ - \mathbf{J}_q^-)/(2 A_{abs})$ where $q$ is the interaction matrix row index, defined as
$q = m \times N_{\text{segments}} + n$, with $m \in \{0,1,2\}$ indexing the mode, $n \in \{0,\ldots,36\}$ denoting the segment number, and $N_{\text{segments}} = 37$ the total number of segments. Each $\mathbf{J}_q$ vector contains $l$ values, corresponding to the number of pixels considered within the MOWFS image. 

The terms $\mathbf{J}_q$ for all the modes and segments are then stored into $\mathbf{J}$. As the resulting matrix is generally non-invertible, a singular value decomposition (SVD) without truncation is used to compute its pseudo-inverse. This yields the command matrix $\mathbf{C}$ which is then used to reconstruct the segment aberrations for a given DM command.

A given measurement vector $\mathbf{b}$ is then produced by recording a raw MOWFS image before applying dark subtraction, normalization by the mean pupil intensity in the absence of the ZWFS mask, subtraction from a reference image acquired with the DM in a flat configuration, and finally vectorization of the resulting two-dimensional image. By multiplying $\mathbf{C}$ with $\mathbf{b}$, the aberration amplitudes are retrieved, yielding a vector of DM commands $\mathbf{v}$ containing the amplitude of each PTT mode for every segment.

For the commands sent to the DM, values of 0 and 10\,nm are considered. We first performed preliminary tests of about 30 and 90\,min, with a single data point every 6\,s, and the evolution trend was seemingly the same. Due to the large amount of data, we decided to limit our run to 30\,min for each mode and segment. However, as a future prospective, it could be informative to analyze data with lower and higher temporal frequencies by performing much longer stability runs. The acquisition rate of one data point every 6\,s with an integration time of 10\,ms was chosen to have a reasonable amount of data to work with over the duration of a run.
We set the DM in flat position and record its motion with the MOWFS to assess the DM stability. We repeat the same operation with a given command leading to 10\,nm for a mode on a given segment of the DM. With the recovered intensities on the MOWFS detector, and the previously calibrated $\mathbf{C}$, we compute the reconstructed commands on the DM. Knowing that the selected segment should be at 0 or 10\,nm, for a specific mode, by comparing with the reconstructed amplitudes, we can estimate the IrisAO DM stability through the temporal variations. 

Based on this acquisition protocol, the temporal stability of the DM is assessed by reconstructing the applied commands from the MOWFS measurements. The recorded intensities are converted into reconstructed amplitudes using the calibrated interaction matrix, allowing a direct comparison between the expected and measured values for each mode and segment. In the following, our analysis focuses on the temporal evolution of these reconstructed amplitudes and on the data noise mitigation to reliably quantify the DM stability.

\subsection{Data noise reduction}

\begin{figure}[ht!]
    \centering
    \includegraphics[width=1\columnwidth]{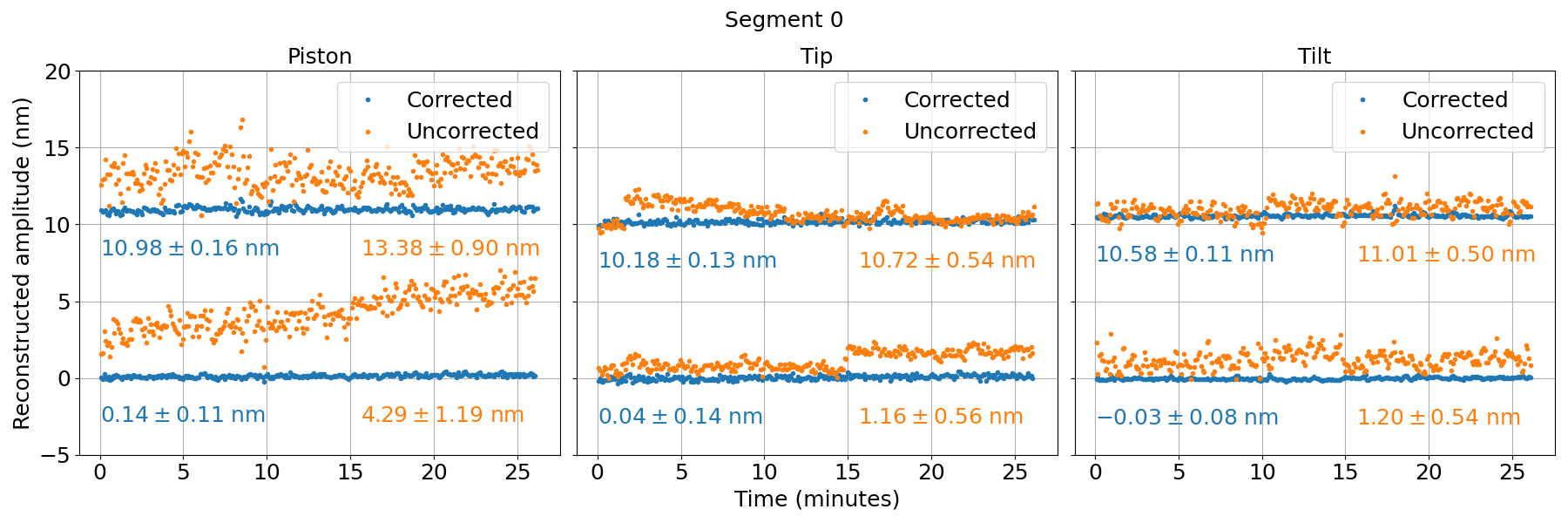}
    \caption{Temporal wavefront error response for a given mode on the central segment with an amplitude of 0 or 10\,nm. For each amplitude case, we compare the uncorrected data affected by turbulence in orange and the data corrected with a synthetic reference signal in blue. From left to right, the plots correspond to the response to a piston, tip, tilt, respectively.}
    \label{fig:wfe_noise}
\end{figure}

Figure~\ref{fig:wfe_noise} represents the temporal evolution of the reconstructed amplitude values on the central segment over 30\,min. Each plot corresponds to a specific aberration mode, piston, tip and tilt. The curves display the response of the measurements for two cases: no command (0\,nm) and a command of 10\,nm PTV sent to the DM. As raw measurements showed some noise, we chose to mitigate it with different strategies of noise reduction. In the following, we present the results with and without noise reduction. The orange points show the data variation without noise correction, and the blue points correspond to the corrected data in post-processing.  

The orange curves indicate that the initial experimental data were too noisy, with average standard deviations going from 0.5\,nm for the tilt to 1.2\,nm for the piston. Instead of curves that should be relatively flat with only a small level of variations, the data seemed to diverge with time. The raw curve show jumps, as the orange curve for the piston mode in Figure \ref{fig:wfe_noise} around a poke value of 0 and 10\,nm. Through further investigation, we found a few sources of errors that needed to be addressed in order to obtain exploitable data.

HiCAT has some known internal air turbulence\cite{Pourcelot_2023}. Additional turbulence is generated by the flip mount motor of the pick-off mirror sending the beam from the HiCAT main optical path to the MOWFS setup. The flip mount becomes hot at use and introduces perturbations as it is located under the beam. Custom modified flip mounts without this issue are being made in-house but are not yet available. The combination of both air and flip mount-based turbulence contributions induces drifts in the measurements, with the mount turbulence seemingly predominant. We note that the piston appears more affected by the turbulence than the other modes. This phenomenon can be explained if we consider being within the framework of the Kolmogorov turbulence regime. 
Piston, tip and tilt correspond to the lowest spatial frequencies of the Kolmogorov spectrum, which contain most of the turbulent power\cite{Noll_1976}. By assuming the Kolmogorov regime at the pupil scale, the tip-tilt modes are the most turbulent modes. So locally at the segment scale, this turbulence contribution can be interpreted as piston variations. If we now consider the Kolmogorov regime at the segment scale, piston, tip and tilt are the most turbulent modes. Combining both considerations leads to a stronger contribution of the turbulence over the piston mode than the tip and tilt modes. 

\begin{figure}[ht!]
    \centering
    \includegraphics[width=0.7\columnwidth]{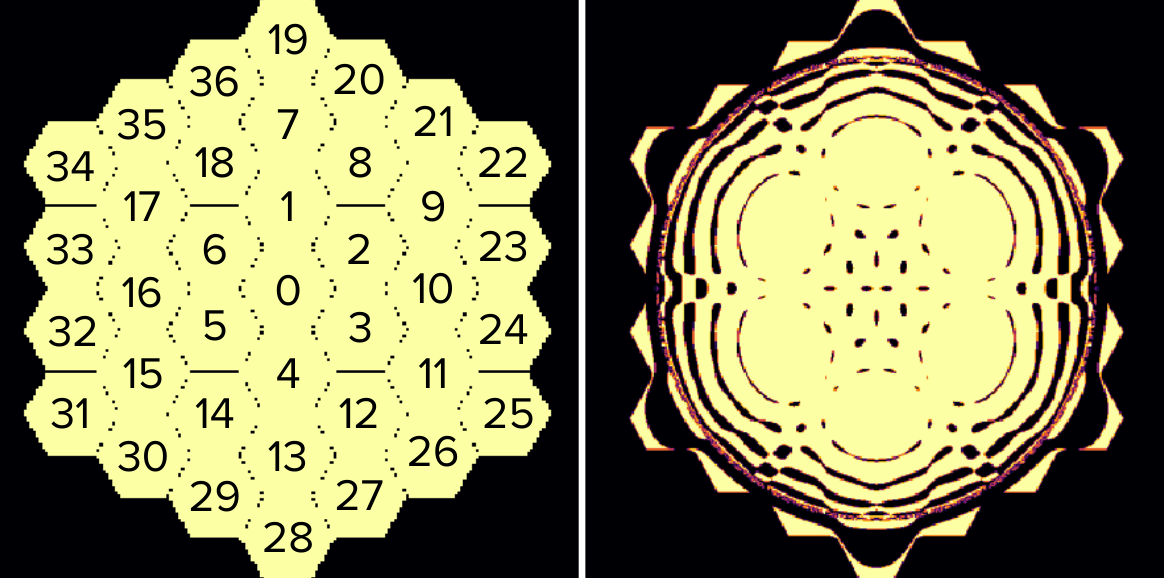}
    \caption{Left: Simulated pupil image of the IrisAO DM surface with the index of the segments. Right: Apodizer shape.}
    \label{fig:apodizer}
\end{figure}

To deal with this source of error in our data, we decided to create a synthetic reference signal that only captures the turbulence contribution. Since the DM is segmented, we can actuate a single segment only while keeping the others steady, and use these stationary segments as a reference.  Using the neighboring segments allows us to sample the local phase under nearly identical turbulence conditions, since most of the Kolmogorov turbulence is in low-order modes with spatial scales larger than this local neighborhood. By averaging the reconstructed amplitudes of these reference segments, we generate a synthetic signal representing the local turbulence. This averaged signal is then subtracted from the reconstructed amplitude of the actuated (poked) segment, effectively removing the estimated turbulence contribution from the measurement, see Figure~\ref{fig:wfe_noise}. For example, when actuating the segment 0, we use the surrounding ring of six adjacent segments (segments 1 to 5, see Figure \ref{fig:apodizer} left frame) to compute the average flat reconstructed commands. We avoided including more distant segments to remain as close as possible within the same turbulence conditions. The limitation of this method is that it performs poorly for the outermost ring of segments, which have fewer neighboring references available.

A second identified source of error is related to the instability of the pick-off mirror in its mount. We observed that when switching from a position to the other, the mirror tends to slightly move in its mount, changing the alignment of the beam on the ZWFS mask. The flip mount specifications give a flip to flip repeatability of 50 $\mu$rad\cite{thorlabs_flip_mounts}, corresponding to 25\,$\mu$m, or 1.5\,$\lambda/D$ on the mask. On the bench, we measured a misalignment of around 8.25\,$\mu$m, which corresponds to 0.5\,$\lambda/D$. Such a drift will alter the measurements of the ZWFS. To compensate for that and ensure a good, stable alignment, we developed a code to automatically finely align the ZWFS mask on the beam. This algorithm is based on the minimization of the global tip-tilt measurement of the intensity on the MOWFS detector. We achieved alignment within a circle of approximately 0.85\,$\mu$m in radius, corresponding to 0.05\,$\lambda/D$. The blue points in Figure \ref{fig:wfe_noise} correspond to the data after applying both turbulence and alignment correction. 

\subsection{Results and discussion}

The blue points of Figure~\ref{fig:wfe_noise} represent the reduced data for the temporal wavefront error (WFE) response for all three PTT modes when amplitudes of 0\,nm and 10\,nm are sent on the DM central segment. It shows some consistency between the introduced and measured WFE for each mode, and we can estimate a temporal evolution for the PTT modes of 5.9, 4.8 and 4.1\,pm RMS/min on the central segment, showing very good stability for a given segment of the DM. 

\begin{figure}[ht!]
    \centering
    \includegraphics[width=0.7\columnwidth]{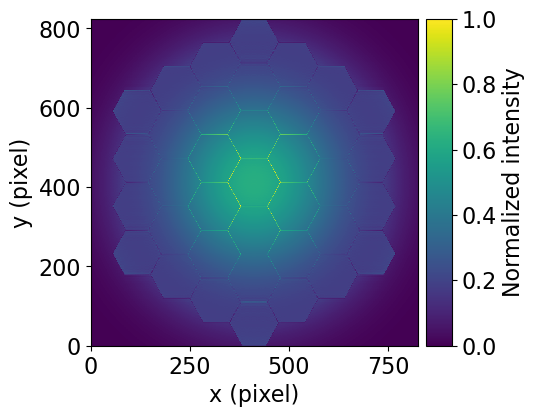}
    \caption{Simulated intensity on the MOWFS detector with the DM in flat position}
    \label{fig:MOWFS_img}
\end{figure}

\begin{figure}[ht!]
    \centering
    \includegraphics[width=1\columnwidth]{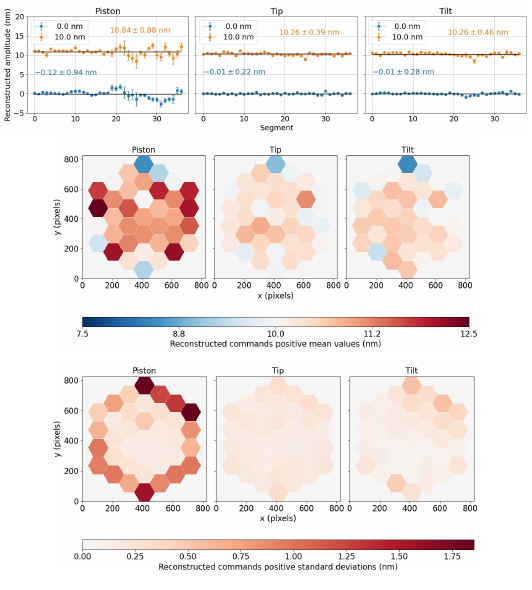}
    \caption{\textbf{Top:} Statistics of the temporal WFE response for each DM segment. The DM stability maps in the middle and bottom plots represent the accuracy and precision of the response of the segmented DM to a given DM command. \textbf{Middle:} Temporal average measurements. \textbf{Bottom:} Temporal standard deviations.} 
    \label{fig:wfe_maps}
\end{figure}

We extend this analysis to the overall segments, see Figure \ref{fig:wfe_maps} top plot. Again, the two cases of 0\,nm and 10\,nm are displayed. It shows a consistency of the response for every mode and each segment. We observe a slight discrepancy of the data for the segments located in the outer rings, mainly for the piston mode. This divergence can be explained by the fact that the outer segments show a less efficient wavefront error correction of the turbulence than the inner segments, because they do not have as many reference segments for the noise data reduction. Another aspect is related to the ZWFS signal. In the absence of errors, the reference intensity on a ZWFS detector in the re-imaged pupil plane exhibits a Gaussian-like shape  (see Figure \ref{fig:MOWFS_img}). This leads to less energy as we go further away from the center, reducing the ZWFS sensitivity for the pixels at the pupil edge which correspond to the signal in the outer segments. It will result in a degradation in temporal WFE standard deviation as we move away from the pupil center, increasing from 0.29 to 1.26\,nm with the segmented mirror in flat position, and from 0.42 to 1.15\,nm with the 10-nm command, in the piston case. Such a behavior may affect the accuracy of the results.

We can estimate the average temporal standard deviation of wavefront errors across all segments and find values of 22$\pm$18, 7$\pm$2 and 7$\pm$5 pm RMS/min for the PTT modes. This result is illustrated by DM stability maps in Figure~\ref{fig:wfe_maps} middle and bottom plots. The maps on the middle and bottom rows correspond to the average reconstructed WFE and the corresponding temporal mean standard deviations. We observe a temporally and spatially uniform response of the segment motion across the full DM. Again, there is more temporal WFE standard deviation with piston and for the external segments, which is most likely due to the imperfect noise reduction in the data. Overall, these results show the excellent temporal stability of wavefront errors due to the segments on HiCAT at level of a few tens of picometers. 

Table \ref{tab:results1} summarizes the stability results for the segments, giving us some insights on the IrisAO DM stability. The turbulence contribution has been dealt with to some extent, but remains an issue for the outer segments. Having a more stable setup would provide more homogeneous stability maps and more accurate information on the turbulence levels. Still, the measurements on the inner segments have been properly denoised for the study of the segmented DM temporal stability. These wavefront error estimates prove very interesting in assessing the impact of the segmented DM behavior in the achievable contrast.

\begin{table}[h!]
\centering
\caption{Comparison between input and measured segment amplitudes for piston, tip, tilt modes sent on central segment and all segments.}
\begin{tabular}{||c||c|c|c||} 

 \hline
 Amplitudes (nm) & Piston & Tip & Tilt \\ 
 \hline
 \hline
 Input & \multicolumn{3}{|c|}{10} \\
 \hline
 Central segment & 10.98 $\pm$ 0.16 & 10.18 $\pm$ 0.13 & 10.58 $\pm$ 0.11 \\ 
 \hline
 Segments 0 to 18 (mean) & 10.99 $\pm$ 0.42 & 10.44 $\pm$ 0.34 & 10.47 $\pm$ 0.23 \\
 \hline
 All segments (mean) & 10.84 $\pm$ 0.88 & 10.26 $\pm$ 0.39 & 10.26 $\pm$ 0.46 \\
  \hline \hline
  Input & \multicolumn{3}{|c|}{0} \\
 \hline
 Central segment & 0.14 $\pm$ 0.11 & 0.04 $\pm$ 0.14 & - 0.03 $\pm$ 0.08 \\ 
 \hline
 Segment 0 to 18 (mean) & 0.16 $\pm$ 0.28 & -0.01 $\pm$ 0.16 & 0.07 $\pm$ 0.13 \\ 
 \hline
 All segments (mean) & -0.12 $\pm$ 0.94 & -0.01 $\pm$ 0.22 & -0.01 $\pm$ 0.28 \\ 
 \hline
 \hline
\end{tabular}
\label{tab:results1}
\end{table}

Our study has so far assumed no correlation between modes and between segments. This hypothesis is based on the fact that each segment is isolated from the others and the actuators under each segment are not correlated in this device. To assess this assumption, we perform a preliminary correlation study using our previous stability measurements. We first investigate the correlation between piston, tip, and tilt on a given segment by sending a 10\,nm command for each mode. We secondly analyze the impact of a given segment mode on all the modes and segments by applying a 10\,nm command on the central segment, assuming a similar behavior for all the other segments. In both cases, the results of our tests show no obvious correlation between the modes of each segment, confirming the validity of our initial assumption. In this analysis, we only considered a command amplitude of 10\,nm. As a future prospect, it would be interesting to perform a more complete correlation study by exploring a wider range of amplitudes to have a more accurate representation of the segmented DM. An additional study would be to look at the presence of electrical correlation that could appear across the actuator channels and possibly impact the segmented DM stability. 

\subsection{MOWFS capabilities}

\begin{figure}[ht!]
    \centering
    \includegraphics[width=1\columnwidth]{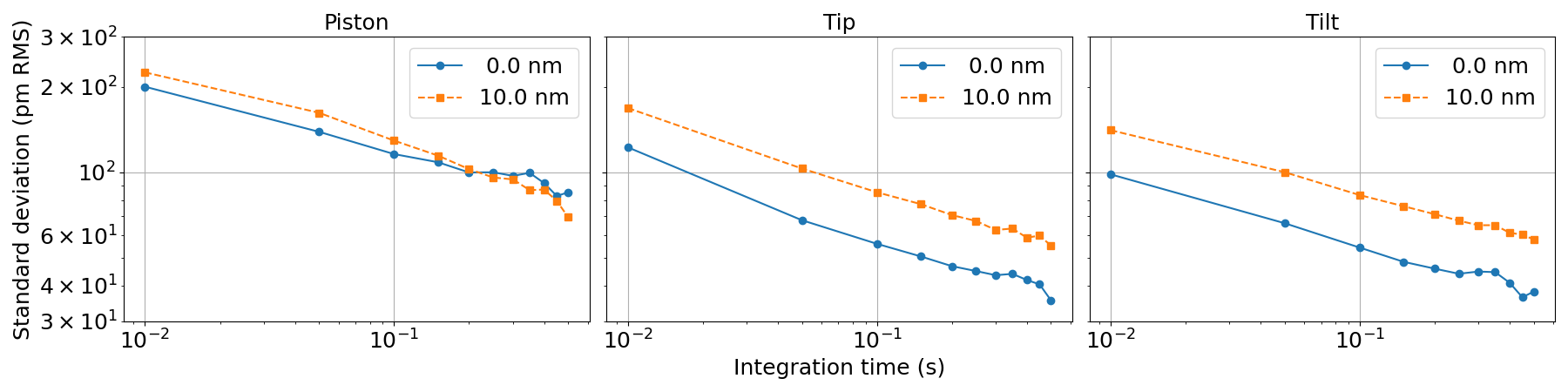}
    \caption{Variation of the averaged WFE standard deviation over the 19 first segments as a function of the integration time, in log scale. Piston, tip and tilt are represented from left to right. The blue solid lines correspond to the flat DM case and the orange dashed lines show the 10\,nm command.}
    \label{fig:wfe_bin}
\end{figure}

With the segment stability study, we also investigated the MOWFS sensing capabilities. Using the corrected temporal variations of the segments 0 to 18, we estimated the precision limit of our sensor by integrating our measurements over different durations with temporal data binning. Figure \ref{fig:wfe_bin} represents the variation of the averaged WFE standard deviation over the 19 segments as a function of the MOWFS integration time. As previously, the flat DM and the 10\,nm cases are represented for the three considered modes. The standard deviation decreases quickly as the integration time increases, seemingly following a power law of about -0.25, which is completely unexpected. Indeed, for an ideal random process, the standard deviation will drop as the $1/\sqrt{N}$ with N denoting the number of samples. At a uniform sample rate, it will drop with $1/\sqrt{t}$ with t representing the time. On a log-log plot, it means a slope of -0.5 instead of -0.25 as observed in our results. Beyond the possible sources of errors, this somewhat unexpected behavior may find its origin in the presence of the aliasing errors due to the discrepancy between the exposure time (10\,ms) and the sampling period (6\,s). This can also be related to the presence of temporally-correlated noise which is exhibited by the computation of the power spectral density (PSD) of wavefront errors reported later in Section \ref{sec:synthetic_wfe}. Further investigation would help disentangling the different sources of error but this is out of the scope of the paper. We here limit ourselves to provide upper limits for the MOWFS capabilities. The curves drop quickly from 10\,ms to around 200\,ms of integration time, before reaching a threshold. In all three cases, the standard deviation reaches values smaller than 100\,pm RMS for all the modes and 60\,pm RMS for the tip and tilt for sensor measurements with an integration time of 200\,ms. While these values represent upper limits in the presence of different error sources, they illustrate the sub-nanometric precision achieved by our MOWFS on HiCAT for the measurements of segment errors.

\section{Contrast limit estimates with the segmented DM on HiCAT}
 
Our goal here is twofold: (i) we want to exploit these WFE stability results to estimate the impact of the segmented DM on the dark zone contrast in the coronagraphic images and (ii) we wish to determine if the WFE variations due to the DM represent a current limitation to achieve deeper contrasts on HiCAT. This analysis will let us know whether it is possible to produce dark hole with deeper contrast with the current IrisAO DM, or the device needs to be improved or changed. 

\subsection{Assumptions}

This study is based on simulation results, using the digital twin of HiCAT. It aims to quantify the contribution of DM stability to contrast degradation, through comparison between simulations and experimental on-bench results from the 2024 HiCAT paper. 

The numerical simulation was conducted with the testbed operating in APLC mode. The APLC suppresses stellar diffraction by combining three optical elements: a pupil apodizer that smooths the amplitude of the electric field, a focal plane mask that blocks on-axis starlight, and a Lyot stop that removes residual diffracted light in the relayed pupil plane, thereby enhancing the detectability of nearby faint sources. The segmented aperture and the apodizer are displayed in Figure~\ref{fig:apodizer}.

On-bench experiments were performed in narrowband (3\%) using three sensing wavelengths: 660\,nm, 680\,nm, and 700\,nm. The dark zone extends from an inner working angle (IWA) of 4.4\,$\lambda/D_{\mathrm{Lyot}}$ to an outer working angle (OWA) of 11\,$\lambda/D_{\mathrm{Lyot}}$, where $D_{\mathrm{Lyot}}$ is the Lyot stop diameter ($D_{\mathrm{Lyot}} = 15$\,mm). The typical measured contrast in this configuration was $2.5 \times 10^{-8}$.

\subsection{Simulation procedure}

Since the objective is to estimate the impact of specific measured variations on the resulting science images, a perfectly flat DM model was used as a reference. This approach isolates the effects of DM stability from any residual static aberrations, while also enabling the development of a more realistic DM model that reproduces the physical behavior of the IrisAO device. First, we have to generate the best achievable contrast on the simulator, which is expected to be below $10^{-8}$.

To reach our contrast goal, we have to run an EFC procedure in the simulator to create the high-contrast region in the coronagraphic image. The EFC algorithm creates a dark hole (DH) by using a pair-wise estimation for the sensing of the electric field with single-actuator probes, through the use of a combination of the two BMC DMs \cite{Borde_2006, Giveon_2007, Will_2023}. When the EFC code achieves convergence towards a flat regime with a satisfying contrast, we stop the procedure and save the last BMC DMs commands, reaching a contrast of 0.5$\times$10$^{-8}$. We then apply these commands to the BMC DMs to reproduce the same dark hole.

Building on the experimental DM wavefront error (WFE) measurements presented in the previous section, we now investigate their impact on coronagraphic performance through numerical simulations. These measurements are used to generate temporal commands applied to a simulated DM, reproducing its physical variations over time. This approach establishes the link between the experimentally characterized DM stability and the corresponding effects on achievable contrast levels.

\subsection{Synthetic WFE: generation and impact}
\label{sec:synthetic_wfe}

\begin{figure}[ht!]
    \centering
    \includegraphics[width=0.8\columnwidth]{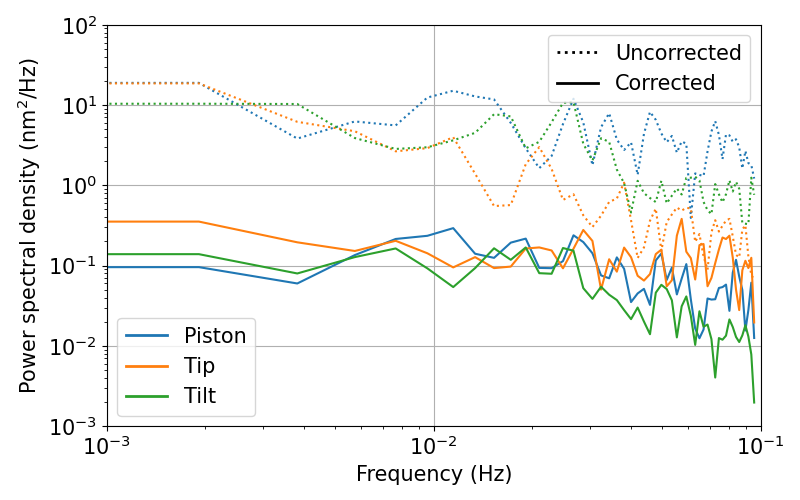}
    \caption{Power spectral density computed with the commands reconstructed from the MOWFS detector response to piston, tip and tilt aberrations on the central segment. The dotted and solid lines show the PSDs of the uncorrected and corrected commands respectively. Both axis are in log scale.}
    \label{fig:PSD}
\end{figure}

In Section \ref{sec:DM_stability}, we showed that the measurements exhibit larger fluctuations from the central segment toward the outer ring, due to less effective turbulence correction and lower signal-to-noise ratio (SNR). For the injection of disturbances of the segmented DM, we chose to base our model on the statistics of the central segment, as it was the most effectively denoised. For all the segments, we individually introduce a behavior with the same statistics as the central one. We computed the temporal PSD of the wavefront errors for the PTT modes, see Figure \ref{fig:PSD}. The PSDs were computed using the Welch's method with a Hann window (size of 100 points per data segment for our dataset of 299 points) to smooth the corresponding curves. From the PSDs of the corrected commands, we generated a synthetic random signal that reproduces the statistical behavior of the central segment.

\begin{figure}[ht!]
    \centering
    \includegraphics[width=1\columnwidth]{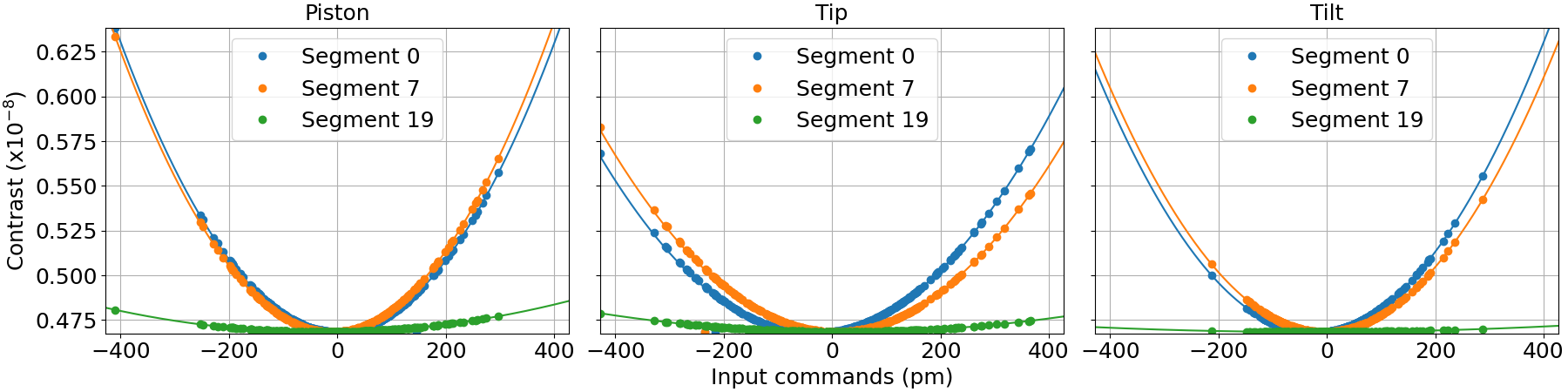}
    \caption{Contrast response curves for different poked segments for all PTT modes. Segment 0 in blue is the central one, segment 7 in orange is within the second ring and segment 19 in green is within the third ring (see Figure~\ref{fig:apodizer}). These plots show the low impact of the outer segments aberrations on the contrast.}
    \label{fig:sensitivity}
\end{figure}

We also examined how the response in contrast varies with the DM ring of the considered segment. Figure~\ref{fig:sensitivity} shows results when we poke one specific segment, on different rings. The data are fitted by a parabolic function. Since the contrast is known to be proportional to the variance of the wavefront errors \cite{Crossfield_2007}, the response curves with a parabolic shape is consistent with our expectations.

These curves shows that for the central segment and the second ring, the response in contrast are similar. However, for the outer ring, the response is weaker. It seems to be mostly due to the apodizer that hides part of the outer ring segments (see Figure~\ref{fig:apodizer}). It is also due to the Lyot stop that is slightly smaller than the pupil and cuts part of the outer segments. Also, as the apodizer may not be perfectly located in the pupil plane, this misalignment may introduce Fresnel propagation effects leading to asymmetric wavefront errors between positive and negative terms, which may be the cause of the curves asymmetry observed in Figure~\ref{fig:sensitivity}.

\subsection{Results}

\begin{figure}[ht!]
    \centering
    \includegraphics[width=0.9\columnwidth]{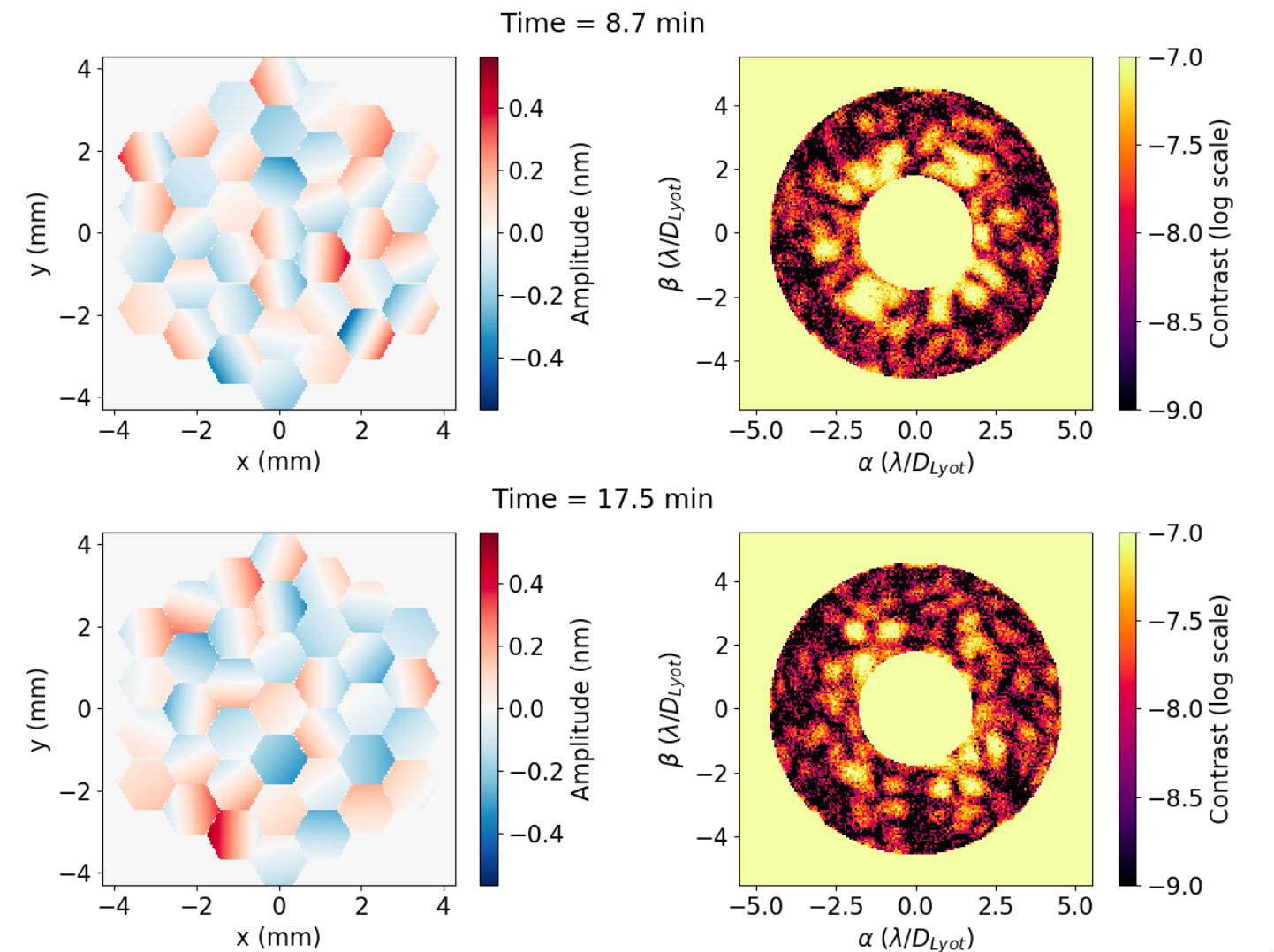}
    \caption{Impact of segments aberrations on the coronagraphic image high contrast zone. Left: Aberrations amplitudes on the IrisAO DM surface. Right: Coronagraphic image high contrast region.}
    \label{fig:maps}
\end{figure}

We consider the case in which all segment and modes are combined, using the synthetic temporal signal derived from the measured variations of the central segment. Figure~\ref{fig:maps} shows examples of two snapshots of the IrisAO DM surface shape and the corresponding high-contrast regions at specific time steps.

\begin{figure}[ht!]
    \centering
    \includegraphics[width=0.8\columnwidth]{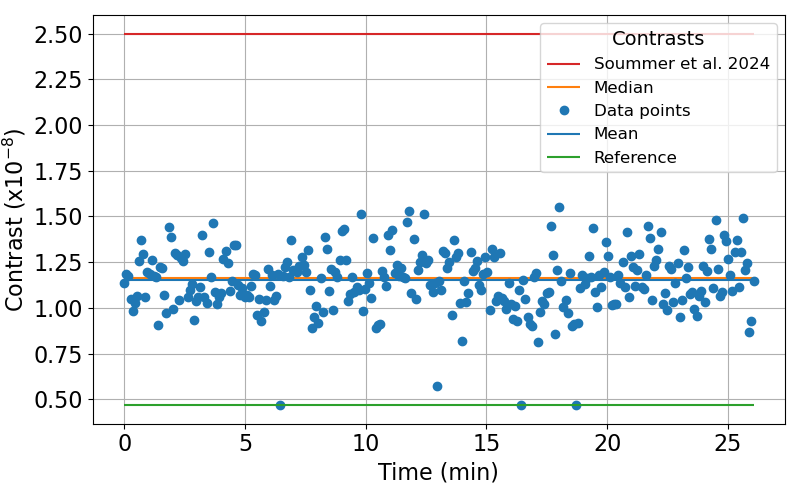}
    \caption{Evolution of the coronagraphic image dark zone contrast as a function of time. The blue dots represent the mean computed contrast values over the high contrast region at a given time. The green line corresponds to the dark zone static flat segmented DM contrast, with no segment misalignment. The red line shows the contrast measured in the paper \cite{Soummer_2024}}
    \label{fig:contrasts}
\end{figure}

Figure~\ref{fig:contrasts} shows the temporal evolution of the coronagraphic dark zone contrast over 30\,min, obtained by injecting a synthetic temporal signal derived from experimental segment stability measurements to the DM. For comparison, the static flat segmented DM simulation contrast is shown as a green line. The red line corresponds to an experimental reference contrast previously reported in the 2024 HiCAT paper. The latter value serves as an experimental benchmark for the analysis.

The data points exhibit a contrast of $(1.15 \pm 0.16)\times10^{-8}$, corresponding to a factor of 2.5 with respect to the contrast with the segmented DM in static flat position. Compared to the contrast values with the central segment only, this is much higher, which shows the high impact of the combination of segments and modes on the DM. The contrast measured and presented in the 2024 HiCAT paper of $2.50\times10^{-8}$ shows a factor of 2.2 with respect to the simulation results. Table \ref{tab:results2} summarizes the different results of our simulation.

\begin{table}[h!]
\centering
\caption{Comparison of achieved contrast in simulation based on measured DM stability with flat DM and experimental contrasts.}
\begin{tabular}{||c||c||} 
 \hline 
 Mean & (1.2 $\pm$ 0.2)$\times 10^{-8}$ \\ 
 \hline
 Median & 1.2 $\times 10^{-8}$ \\
 \hline
 Static flat segmented DM & 0.47 $\times 10^{-8}$ \\
 \hline
 2024 HiCAT paper & 2.5 $\times 10^{-8}$ \\
 \hline
 \hline
\end{tabular}
\label{tab:results2}
\end{table}

\subsection{Discussion}

Through this experiment, we wanted to characterize the impact of the DM alignment stability on the science image contrast. The idea was to determine to what degree the behavior of the DM can explain the difference between the simulated dark hole contrast and the experimental one. Figure~\ref{fig:contrasts} shows a factor of 5.32 in terms of contrast between the segmented DM simulated in static flat position and the experiment from the 2024 HiCAT paper. Our simulation gave us an intermediate result between the two reference values. We can conclude that the whole contrast degradation can be partly explained by adding the measured variation of the DM surface in the simulation. The degradation may also be caused by other elements on the testbed that we did not consider here. For instance, there might be stability issues on other active components, such as the BMC DMs. The turbulence contribution may also affect other optical path on the testbed, degrading the contrast limit. 
We also know that the segmented DM numerical model is not complete. In a previous study \cite{Buralli_2024}, we presented preliminary experiment results showing a non-linear response of the DM at very low amplitudes through the presence of discretization steps. The combination of all these factors could explain the gap between the measured and simulated contrasts.

\section{Conclusion}

We presented the results focused on the impact of a segmented DM stability on image contrast on HiCAT. The experimental results were obtained using a ZWFS installed on HiCAT, with the goal of evaluating the influence of the DM segment misalignment on the contrast. This experiment allowed us to assess the DM stability for each of the 37 segments and all three PTT modes. We observed good consistency between the introduced and measured WFE for each of the PTT modes, and derived an estimate of the temporal evolution for those modes: 5.9, 4.8 and 4.1\,pm RMS/min, on the central segment. From these data, we also determined average values of the temporal standard deviations over a 30-min range: 22 $\pm$ 18, 7 $\pm$ 2 and 7 $\pm$ 5\,pm RMS/min for piston, tip, and tilt, respectively. The segment stability analysis also allows us to characterize the MOWFS performance, showing sub-nanometric precision with WFE standard deviations below 100\,pm RMS for all modes and 60\,pm RMS for tip and tilt with a 200\,ms integration time.

We then used these experimental results to create a numerical model of the segmented DM stability in the HiCAT simulator. Through these simulations, we could study the impact of the DM stability on the dark zone contrast, and observed a degradation by a factor of 2.5, from 4.7$\times$10$^{-9}$ with the DM in static flat position to 1.2$\times$10$^{-8}$ when introducing the stability measurement. By comparing this with the experimental reference contrast of 2.5$\times$10$^{-9}$, we concluded that DM stability can have a significant impact on achievable performance. On HiCAT, we have provided an upper bound of the contribution due to the segmented DM on the contrast degradation, explaining part of the discrepancy between the simulated and measured contrasts. In a previous paper \cite{Buralli_2024}, we reported the presence of quantization steps at low amplitudes. This non-linearity may need to be added to the numerical model to further reduce the contrast discrepancy.

Overall, this work demonstrates the importance of characterizing the temporal stability of segmented DMs and their influence on high-contrast imaging testbeds. Beyond the specific case of the IrisAO device, these results illustrate the broader necessity of accurately modeling and validating segmented mirror behavior for contrast levels at levels of $10^{-9}$ to $10^{-10}$. Such knowledge will be essential to establish robust specifications for future segmented DMs and to ensure that testbeds can realistically emulate the conditions of large space telescopes. Ultimately, improving our understanding of DM stability and possible correlations between modes of each segment directly supports the development of system-level approaches required by flagship missions such as HWO, where the detection and characterization of Earth analogs around Sun-like stars will rely critically on the stability and control of segmented apertures.

\appendix    % this command starts appendixes

\subsection*{Disclosures}

The authors declare that there are no financial interests, commercial affiliations, or other potential conflicts of interest that could have influenced the objectivity of this research or the writing of this paper.

\subsection*{Code, Data and Materials Availability}

The codes and data used in this study are available from the corresponding author upon reasonable request.

\subsection*{Acknowledgments}

The authors thank Damien Sucher for stimulating discussions about the topic. B.B. acknowledges PhD scholarship funding from Région
Provence-Alpes-Côte d’Azur and Thales Alenia Space.
B.B. thanks the Observatoire de la Côte d'Azur for its support. This work was supported by the Action Spécifique Haute Résolution Angulaire (ASHRA) of CNRS/INSU co-funded by CNES. This work was also supported by the TARPIN international research project of the CNRS and laboratoire Lagrange through the 2023 and 2024 BQR Lagrange program. The HiCAT testbed has been developed over the past 10 years and benefited from the work of an extended collaboration of over 50 people. This work was supported in part by the National Aeronautics and Space Administration under Grant 80NSSC19K0120 issued through the Strategic Astrophysics Technology/Technology Demonstration for Exo-planet Missions Program (SAT-TDEM; PI: R. Soummer), and under Grant 80NSSC22K0372 issued through the Astrophysics Research and Analysis Program (APRA; PI: L. Pueyo). E.H.P. was supported in part by the NASA Hubble Fellowship grant HST-HF2-51467.001-A awarded by the Space Telescope Science Institute, which is operated by the Association of Universities for Research in Astronomy, Incorporated, under NASA contract NAS5-26555. Sarah Steiger acknowledges support by STScI Postdoctoral Fellowship and Iva Laginja acknowledges partial support from a postdoctoral fellowship issued by the Centre National d’Etudes Spatiales (CNES) in France.
This research made use of HCIPy, an open-source object-oriented framework written in Python for performing end-to-end simulations of high-contrast imaging instruments (Por et al. 2018 \cite{Por_2018}). This research made use of \texttt{CATKit2}, an open-source package for controlling testbed hardware (Por et al. 2024 \cite{Por_2024}).

%%%%% References %%%%%

\bibliography{report}   % bibliography data in report.bib
\bibliographystyle{spiejour}   % makes bibtex use spiejour.bst

%%%%% Biographies of authors %%%%%

\vspace{2ex}\noindent\textbf{Benjamin Buralli}

Benjamin Buralli is currently a third year PhD student at Lagrange Laboratory in Nice, France. He is working under the supervision of Mamadou N'Diaye and Marcel Carbillet on active control of mid-order aberrations based on Zernike wavefront sensors for exo-Earth imaging with future large space observatories. His research work is performed in collaboration with the Space Telescope Science Institute in Baltimore, USA and Thales Alenia Space in Cannes, France.

\vspace{1ex}
\noindent Biographies and photographs of the other authors are not available.

\listoffigures
\listoftables

\end{spacing}
\end{document}